\documentclass{aa}  
\usepackage{txfonts,textcomp}
\usepackage{appendix}
\makeatletter
\renewcommand*\aa@pageof{, page \thepage{} of \pageref*{LastPage}}
\makeatother
\begin{document}
\title{Forgotten treasures in the HST/FOC UV imaging polarimetric archives of active galactic nuclei}
\subtitle{II. Mrk~463E}
\author{T. Barnouin\inst{1}\thanks{\href{mailto:thibault.barnouin@astro.unistra.fr}{thibault.barnouin@astro.unistra.fr}}
    \and
    F. Marin\inst{1}
    \and
    E. Lopez-Rodriguez\inst{2}
}
\institute{Universit\'e de Strasbourg, CNRS, Observatoire astronomique de Strasbourg, UMR 7550, F-67000 Strasbourg, France
    \and
    Kavli Institute for Particle Astrophysics \& Cosmology (KIPAC), Stanford University, Stanford, CA 94305, USA
}
\date{Received Month Day, 2024; accepted Month Day, 2024}
%
%
\abstract
{The Mrk~463 system is known to host two powerful sources separated by about 4~kpc, both identified as active galactic nuclei (AGN). This makes the Mrk~463 system a unique laboratory to study the geometry and dynamics of galaxy merging and its relation to AGN duty cycles.}
{The eastern nuclei, Mrk~463E, is the brightest of the two and thus a prime target for a polarimetric study. It is classified as a Seyfert~2 galaxy, meaning that one could expect large polarization degrees from scattering off electrons and dust in the polar winds.}
{In the continuity of our series of papers, we reduced archived and previously unpublished polarization observations obtained with the Faint Object Camera (FOC) onboard the Hubble Space Telescope (HST), to obtain a high resolution near ultraviolet (NUV) polarization map of the Mrk~463E nuclei. We coupled this map to near infrared (NIR) and X-ray observations to get a clear picture of the geometric arrangement of matter around the core of Mrk~463E.}
{We found that the nucleus location is further South from the optical peak flux than previously estimated. The strongly polarized conical wind has a half-opening angle of $\sim15^\circ$ and display three main periods of mass ejection. Its polarization allowed us to estimate the AGN inclination towards the observer ($\sim55^\circ$). Finally, our maps revealed a gas streamer connecting Mrk~463E and Mrk~463W, with a tentative detection of a large kpc-scale ordered magnetic field connecting both galaxies.}
{This unpublished observation turned out to offer more than the original proposal asked for and allowed to derive tight geometric and dynamical constraints for Mrk~463E. High resolution radio maps and IR polarimetry are now necessary to further study the jet and the newly discovered gas streamer.}
\maketitle
\section{Introduction}
It is now quite accepted that massive galaxies should host a supermassive black hole (SMBH) at their center. The most recent observations from the Event Horizon Telescope provided powerful evidence for their existence in the nearby galaxy M87 \citep{EHTM872019}, as well as in the center of our own galaxy \citep{EHTSagA2022}. However their formation and growth remains difficult to trace, even more so if we take into account the parallel evolution of their host galaxy \citep{DiMatteo2005}. Massive black hole mergers and powerful accretion phases, responsible for the so-called Active Galactic Nuclei (AGN), are the main processes thanks to which we explain SMBH growth \citep{DiMatteo2005,Treister2012}. However, if taken separately, these growth models remain too slow for both galactic and cosmological simulations to explain the observed SMBH masses with respect to their host galaxy age \citep{Capelo2017,Volonteri2022}. Both processes should occur simultaneously in the lifetime of a black hole to become supermassive. The best place to look for concurrent merging and accretion are probably the Ultra Luminous Infrared Galaxies \citep[ULIRG,][]{Sanders1988}, which are believed to be the ongoing phase or the remnant of a merging process that is able to bring a lot of dust and gas nearby their central SMBH \citep{Barnes1991,Hopkins2006}. This could trigger an AGN phase and lead to intense SMBH growth at the same time \citep{Treister2012}.

Mrk~463 is a nearby, $\sim224$~Mpc (in standard $\Lambda$CDM cosmology), merging system consisting of two nuclei, Mrk~463E (east, $z = 0.05132$, \citealt{Lavaux2011}) and Mrk~463W (west, $z = 0.05055$, \citealt{Abazajian2009}), separated by $3.8'' \pm 0.01''$, which corresponds to a projected distance of $\sim4.1$~kpc. Observations in radio \citep{Neff1988,Mazzarella1991}, [OIII] and [NII] bands \citep{Hutchings1989,Chatzichristou1995} were used to study the interconnected nuclei and revealed tidal tails from a recent violent interaction as well as evidences for the ongoing merging of the two galaxies \citep{Hutchings1989, Mazzarella1991}. High resolution 6~cm radio observations of the eastern nuclei, Mrk~463E, revealed a small-scale jet ($0.05''$ to $1.5''$) along a Position Angle (PA) of $\sim11^\circ$ \citep{Neff1988}. At 20~cm, \cite{Mazzarella1991} observed extended but isolated emission sources situated $18''$ ($\sim20$ kpc) south and $4''$ ($\sim5$ kpc) north of Mrk~463E, without optical or IR counterparts. They are not connected to the bright radio emission of the nucleus but, since they are located along the same axis as the small radio jets and have characteristics of classic radio lobes, they were associated to Mrk~463E as weak radio lobes. The jets associated to Mrk~463E can be extended up to another emission region as far as $\sim12$~kpc to the northwest of the nuclei \citep{Neff1988,Mazzarella1991}. Such extended radio emission and its strong luminosity density put Mrk~463E in-between the Seyfert and radio-galaxy classification.

Due to its early identification as a type 2 Seyfert galaxy \citep{Shuder1981}, the eastern nucleus of Mrk~463 has been the focus of polarization studies to determine if the AGN fits the Unified Model \citep{Antonucci1993}. Indeed, \cite{Miller1990} discovered broad emission lines in its polarized spectrum, proving the presence of a type-1 nucleus behind some obscuring material. Broad (full width half-maximum $\sim2750$ km$\cdot$s$^{-1}$) H$\beta$ and H$\alpha$ Balmer lines were detected in polarized flux, with an intrinsic polarization degree of $10\%$ and an electric vector polarization angle of $85^\circ$ (after correction for interstellar polarization), which was comparable to the other archetypal Seyfert-2 AGN : NGC~1068 \citep{Antonucci1985,Tran1995a,Tran1995c}. \cite{Miller1990} also found a strong 5500~\AA ~continuum linear polarization of $7.7\%$ with a polarization angle of $\sim82^\circ$, almost perpendicular to the radio PA ($\sim4^\circ$, \citet{Unger1986} and\citealt{Neff1988}). Similarly to NGC~1068 \citep{Antonucci1985}, Mrk~463E ionization cones spread along the radio axis and their illumination by the central engine can easily explain such polarization levels and angles. However, the detected narrowing of the [OIII] and [OII] emission lines in the polarized spectrum hinted for unresolved geometric or dynamic features \citep{Miller1990, Tran1995c} and sparkled new studies of Mrk~463E at higher spatial resolutions.

High resolution imaging of the eastern nucleus in the optical band was obtained with the Planetary Camera (PC) onboard the Hubble Space Telescope (HST) in July 1991. This observation led to the discovery of a sub-arcsec scale jet-like structure around the nucleus \citep{Uomoto1993}. While similar in morphology to the optical jets of M87 and 3C~273, the radio emission in Mrk~463E is positioned just beyond the end of this optical "jet". The optical and radio emission do not correlate and therefore the observed optical emission from this "jet" can not be synchrotron in nature. Because the axis of this polar structure is almost perpendicular to the detected continuum polarization, \cite{Uomoto1993} suggested that it was the scattering site that allows the observation of the hidden Seyfert-1, producing the polarized spectrum observed by \cite{Miller1990}. A follow up observation using the Faint Object Camera (FOC) onboard HST was performed in May 1996 to verify this prediction. To this day, those high quality data were never reduced, analyzed nor published, despite some preliminary announcements by \cite{Tremonti1996}.

The study of the Mrk~463 system continued at higher energies, especially because the spectral type and true AGN nature of the western nuclei were never confirmed \citep{Shuder1981,Neff1988}. Chandra high-energy (2--8~keV) imaging analysis of the Mrk~463 system revealed the presence of two unresolved X-ray sources, coincident with the two nuclei detected in radio and IR \citep{Bianchi2008}. X-ray spectral analysis using Chandra, XMM-Newton and NuSTAR data showed that Mrk~463W is a Compton-thin obscured AGN (n$_{\rm H} \le$ $10^{24}$ at/cm$^2$) with a small bolometric to X-ray luminosity ratio ($\kappa_{2-10keV} \sim 1-3$), suggesting a low Eddington ratio \citep[$<10^{-3}$;][]{Yamada2018}. Hence Mrk~463W is most probably in an early phase of AGN activity triggered by the merger event. Regarding the eastern nucleus, the same analysis showed that its intrinsic spectrum is X-ray weak (with respect to its UV luminosity) but is associated with a high Eddington ratio  \citep[$\lambda_\text{Edd} = 0.4-0.8$;][]{Bianchi2008,Yamada2018}. This would suggest a Compton-thin obscured AGN with a rapidly growing SMBH. The recent merger event could have triggered efficient gas fueling onto the SMBH without deeply burying the nuclei with gas and dust.

It is thus evident that, thanks to its distance, luminosity and recent history, the Mrk~463 system makes a unique laboratory to study AGN evolution and SMBH growth. A violent interaction between its two nuclei probably triggered their AGN phase, as observed from the radio to the X-rays. Some high resolution archival data still remain to be analyzed to complete and to better understand how the regions interact, what is the nature of the bright, sub-arcsec scale, optical jet-like structure in Mrk~463E and how matter is arranged around this nucleus. Therefore, in this second paper of our series about the archival HST/FOC data, we reduce and analyse for the first time a NUV imaging polarimetry observation of the scattering regions of Mrk~463E, requested in 1995, to verify the prediction of \cite{Uomoto1993}. In Sect.~\ref{Observation}, we present the observation and data reduction techniques. The reduced polarization map is analyzed in details in Sect.~\ref{FOCAnalysis}, making use of the high spatial resolution of the instrument for mapping several specific regions around the nucleus. Sect.~\ref{MultiwavAnalysis} puts this observation into a multi-wavelength context to better highlight the different processes at play and understand the nature of the jet-like structure. The results of this analysis are discussed in Sect.~\ref{Discussion}. Sect.~\ref{Conclusion} summarizes our findings.

\section{Observation and Data Reduction}
\label{Observation}
Mrk~463E was observed by the HST/FOC on May 10, 1996 (program ID 5960). This observation was requested as a follow-up of the previous HST/PC observation analyzed and published in \cite{Uomoto1993}. Its goal was to inspect in NUV light the sub-arcsecond jet-like structure discovered in the optical and determine if it is, indeed, a scattering site. This observation was made in the f/96 mode with the zoom off. It results in pixels of size $0.014'' \times 0.014''$ for a $7'' \times 7''$ Field of View (FoV). It used the F342W filter centered around 3404~\AA, for a total exposure time of 10,748 seconds, 11,300 seconds and 10,748 seconds through the POL0, POL60 and POL120 filters, respectively. The observation products are stored in different flavors of calibration in the MAST/HST legacy archives of the FOC: either raw uncalibrated data, geometrically and exposure corrected data or geometrically, exposure and flatfield corrected data. The data were retrieved and reduced as described in the first paper of our series \citep{Barnouin2023}. We refer the reader to this paper for an in-depth explanation of the reduction steps and definition of the uncertainties (Section 2.7 in the aforementioned paper, $\sigma_I$ and $\sigma_P$) used in the following analysis. All reported polarization degree are corrected for bias \citep{Simmons1985}.

For this observation, the FoV of the FOC instrument is centered on the eastern nucleus, putting the western nucleus just outside of the image. A quick analysis reveals small to no contamination of light coming from Mrk~463W: only part of an extended source can be seen in the bottom-right corner of the instrument FoV (as seen in Fig.~\ref{Fig:MRK463E_FOC_I}). Thanks to the imaging capacities of the HST/FOC, the direct emission from the western nucleus can be spatially distinguished from the eastern nucleus emission and analyzed separately. For the analysis of Mrk~463E in the following, the extended emission coming from Mrk~463W is cropped out of the region of interest.

The long exposure time through each polarized filter allowed us to obtain statistically significant measurements through our reduction pipeline: in the resampled 256 $\times$ 256 pixel resolution of the FOC (Nyquist sampling of the native 512 $\times$ 512 pixel image), and without smoothing, we obtain a flux detection (at $\geq3\sigma_I$) in $33.8\%$ of the pixels, $0.03\%$ of which have a polarization detection ($\geq3\sigma_P$), amounting to 6 pixels. The retrieved data were not deconvoluted with a Richardson-Lucy algorithm -- contrary to what was done in \cite{Uomoto1993} -- as we want to study the diffuse emission from the Narrow Line Region (NLR) and because of the native high signal-to-noise ratio per pixel in the region of interest. In the following section, we thus present an analysis of the polarization map rebinned to pixels of size $0.05'' \times 0.05''$, i.e. $3.5$ times the native pixel size, and smoothed with a bi-dimensional Gaussian filter with a full-width at half-maximum of $0.10''$, i.e. twice the resampled pixel size. Unlike the first paper of this series, the polarization vectors are shown for a confidence level on the polarization degree greater than 99\%. For every measure, the confidence level is computed for two degrees of freedom -- normalized Stokes Q and U fluxes, see Appendix \ref{App:Conf} for details. Integration inside a region of interest or selected aperture is conducted on the Stokes I, Q and U fluxes, before the computation of the polarization degree and angle and bias correction.

\section{Polarization Analysis}
\label{FOCAnalysis}
\subsection{The FOC polarization map}
\label{FOCmap}
\begin{figure*}
    \centering
    \includegraphics[width=1.\textwidth]{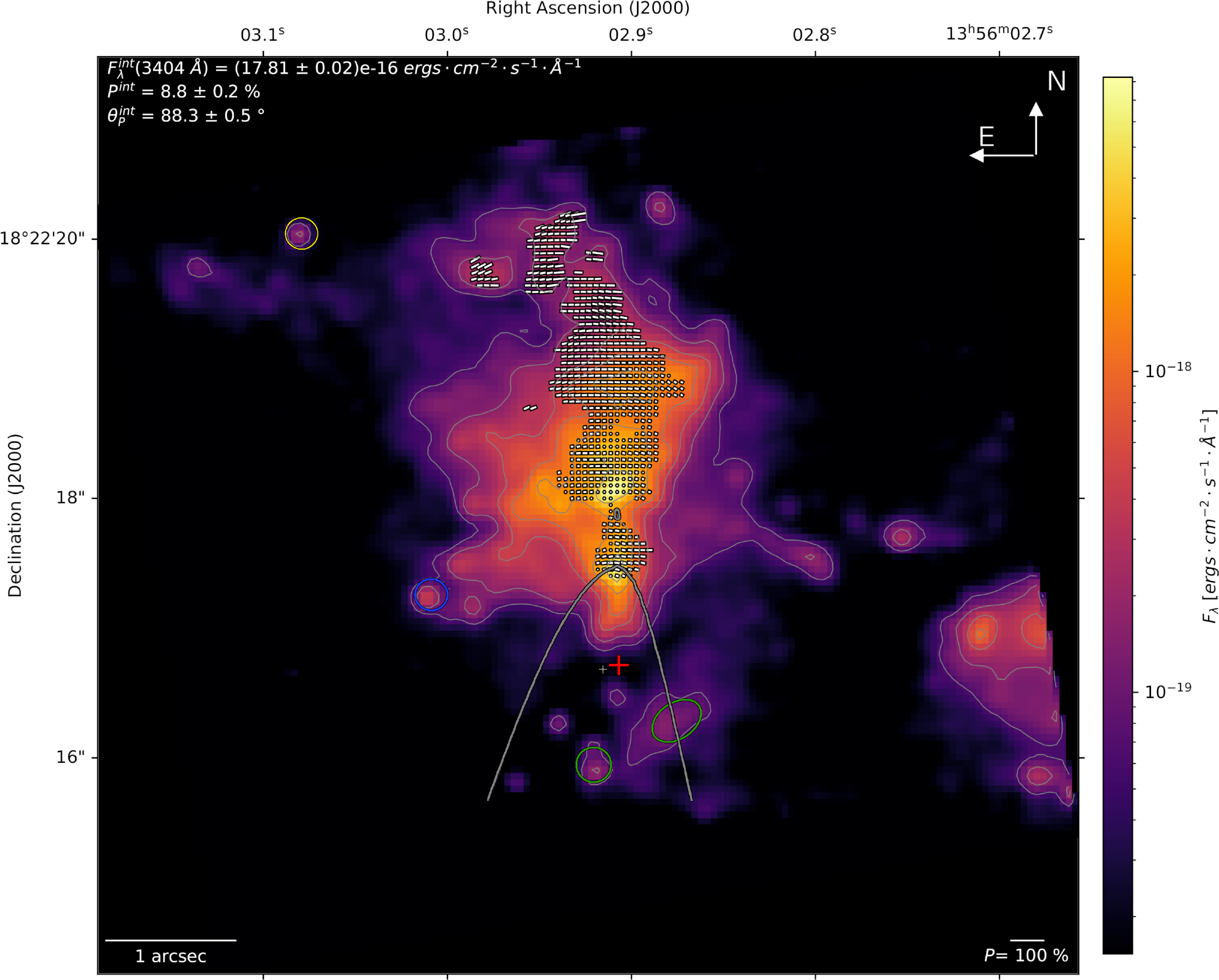}
    \caption{HST/FOC observation of Mrk~463E resampled to pixels of $0.05'' \times 0.05''$. Intensity is color-coded in erg cm$^{-2}$ s$^{-1}$ \AA$^{-1}$. Polarization vectors are displayed for $\text{Conf}_P \geq 99\%$. The dark grey contour and cross, as well as the red cross, show our estimations for the nucleus location and are discussed in Sect. \ref{FOCmap} and Appendix \ref{App:center}. The flux density contours are displayed for 0.8\%, 2\%, 5\%, 10\%, 20\% and 50\% of the maximum flux density. Top-left corner values are integrated over the whole FOC FoV ($7'' \times 7''$): $F^\text{int}_\lambda = (17.81 \pm 0.02) \times 10^{-16}$ ergs cm$^{-2}$ s$^{-1}$ \AA$^{-1}$ with $P^\text{int} = 8.8 \pm 0.2 \%$ and $\Psi^\text{int} = 88.3 \pm 0.5 ^\circ$. Circled spots are discussed in sect. \ref{FOCmap}.}
    \label{Fig:MRK463E_FOC_I}
\end{figure*}

Fig.~\ref{Fig:MRK463E_FOC_I} shows the polarization map overlaid on the total intensity image. The polarization vectors are displayed for $\left[\text{S/N}\right]_P \geq 3$, their length is proportional to their polarization degree and they are oriented according to their electric vector polarization angle, from North to East direction, following the IAU convention \citep{IAU1974}. The total flux density image is dominated by two hot spots of flux density on a North-South axis and whose centers are separated by $\sim0.67$~kpc (projected distance). These large hot spots are connected by a narrow flux density column, as if an external constraint was creating a physical bottleneck between the two extended flux density spots. Intense emission surrounds those two hot spots that forms a well-contrasted, large-scale, cone-shaped region, a geometry already observed for many other Seyfert AGNs \citep[see, e.g.][]{Miller1991,Wang2024,Wilson1996}. Such bi-cone (or hourglass-shaped) structure is known to be associated with the NLR, emerging from the (unresolved) vicinity of the black hole and extending hundreds of parsecs along the polar axis of the AGN, often revealed thanks to [O III] mapping \citep[see, e.g.][]{Walker1968,Oke1968,Mullaney2013}. Indeed, [O III] being a forbidden line, it is able to trace the kinematics in the NLR, which was measured to be of the order of 300 km s$^{-1}$ to 600 km s$^{-1}$ by \citet{Treister2018} in the case of Mrk~463E. The highest velocities are reached in a small region North-East of the northern cone, with an apparent different flowing direction than the NLR itself. In our map, we see that a few ovoid clumps of stronger polarization than the NLR seem to escape the wind towards the North-East direction, strengthening the idea that the NLR is a dynamical region. The southern part of the Mrk~463E's bi-cone is not visible, likely hidden behind the host galaxy plane (see Sec.~\ref{WFPCAnalysis} for further details).

\begin{figure*}
    \centering
    \includegraphics[width=.49\textwidth]{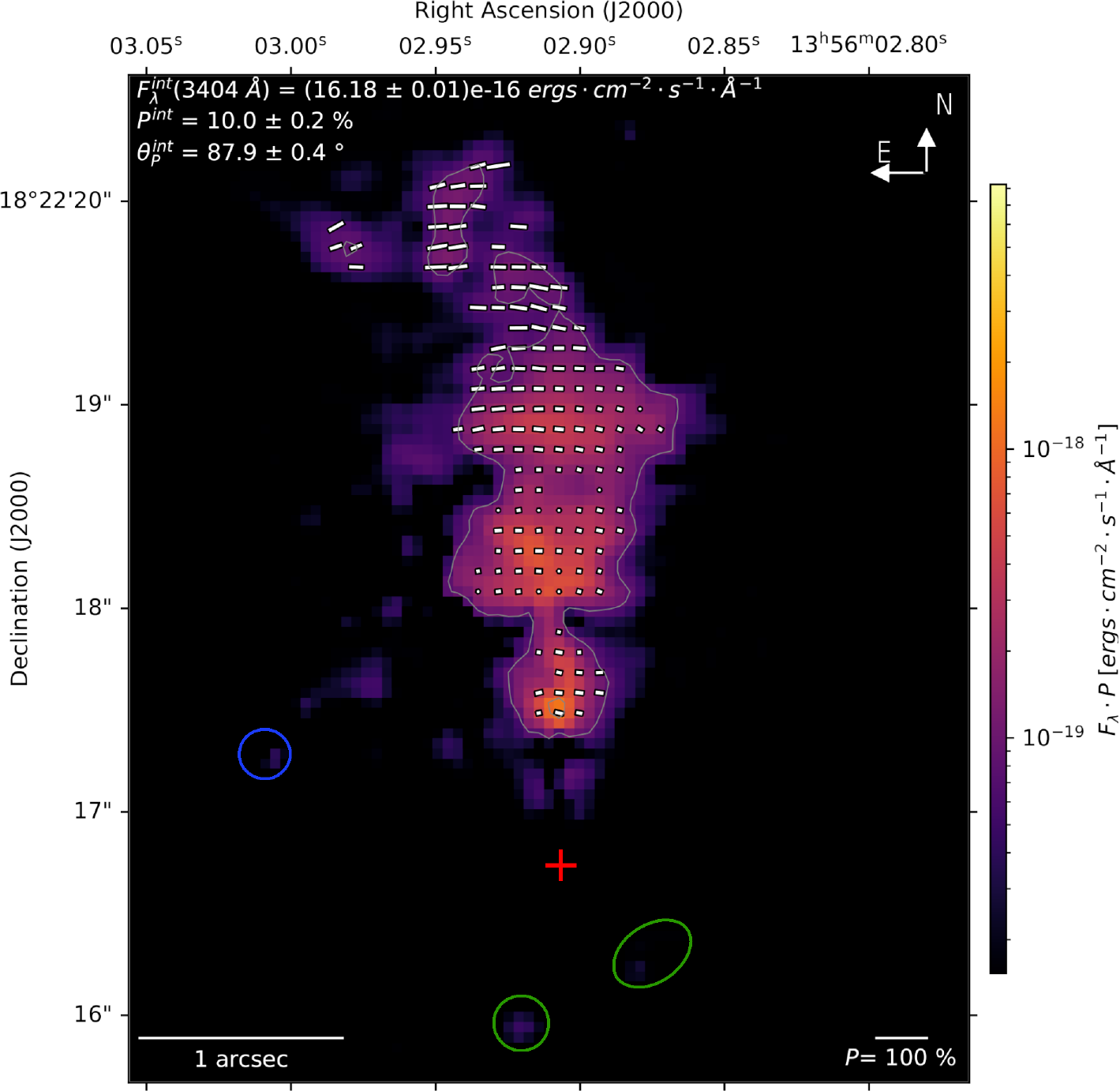}
    \includegraphics[width=.49\textwidth]{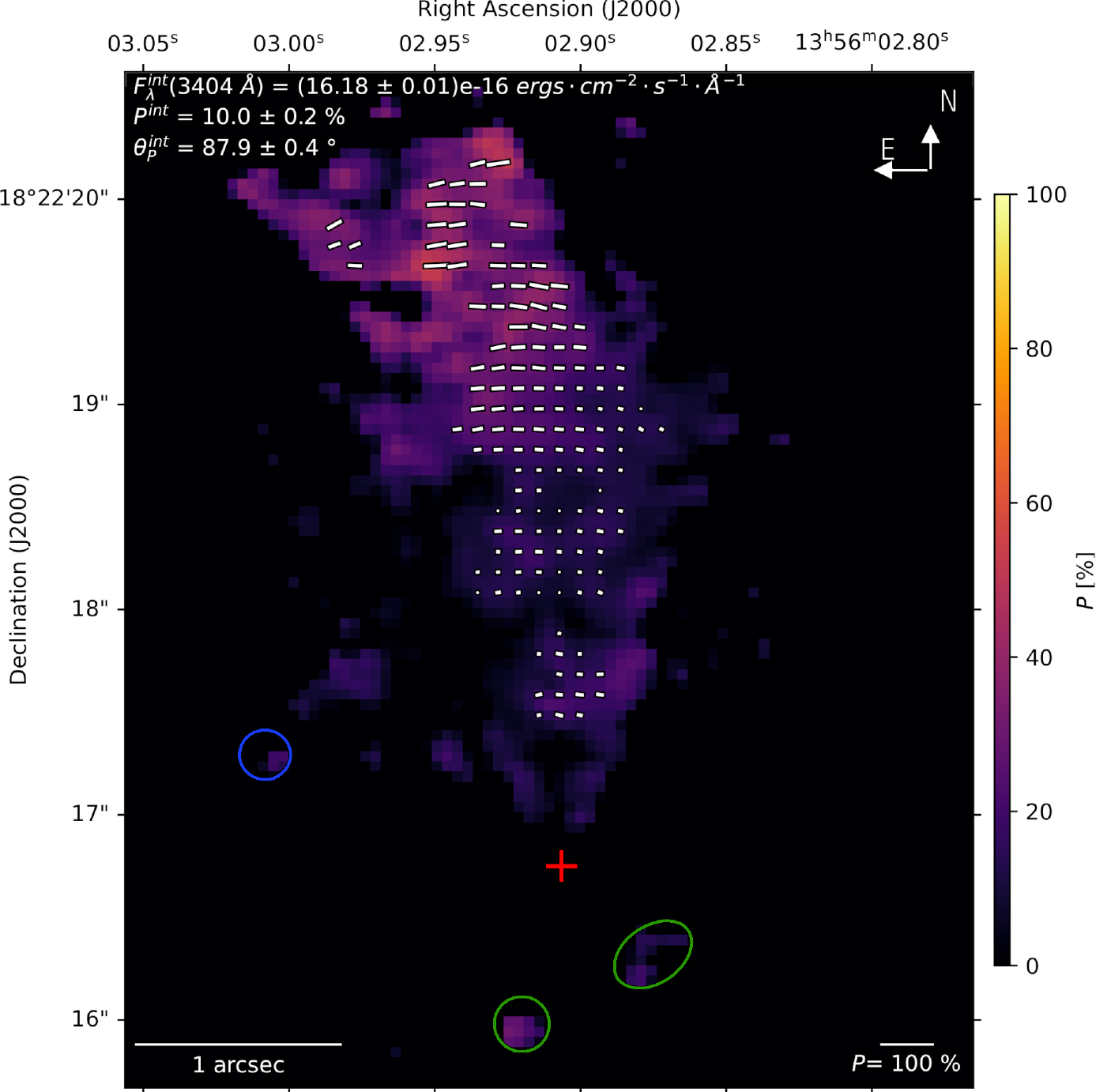}
    \caption{Cropped polarized flux density (left) and polarization degree (right) maps highlighting the northern scattering region of Mrk~463E. The red cross shows our estimation for the nucleus location. Circled spots are discussed in sect. \ref{FOCmap}. For better visibility we only display every other vector.}
    \label{Fig:MRK463E_FOC_PF_P}
\end{figure*}

A second, less extended emission originating from the western nucleus, Mrk~463W, can be seen at the South-West border of the FoV. As we lack a good part of the emission from Mrk~463W, we can not infer the geometrical aspect and emission characteristics of the region appearing in this observation. We report no detectable polarization coming from Mrk~463W: the small polarization features seen in Fig.~\ref{Fig:MRK463E_FOC_PF_P} in the western nucleus remain below $1.5\sigma_P$ and can not be dissociated from noise. However, a stream-like feature extends from the base of the observed polar wind of Mrk~463E in an elliptical arc into the general direction of Mrk~463W. This arc, visible in total intensity (see Fig.~\ref{Fig:MRK463E_FOC_I}), has no $\left[\text{S/N}\right]_P \geq 3$ polarized counterparts (see Fig.~\ref{Fig:MRK463E_FOC_PF_P}). See Sect.~\ref{Filament} for further details about this arc-like structure.

In the polarized flux density map shown in the left-hand part of Fig.~\ref{Fig:MRK463E_FOC_PF_P}, we can see that the aforementioned conic wind and ovoid clumps account for almost all of the polarized flux density. The cone appears much narrower in polarized flux density (half-opening angle of $\sim15^\circ$) that what can be seen in total intensity (half-opening angle of $\sim36^\circ$), both reported on Fig. \ref{Fig:MRK463E_polar}. This is similar to what has been observed in NGC~1068, for which the ionization cone appears much narrower in polarized flux density \citep{Miller1991}. We use both of these cone delimitation to make an estimation of the nucleus location, displayed as a red cross on Fig. \ref{Fig:MRK463E_FOC_I}, \ref{Fig:MRK463E_FOC_PF_P} and \ref{Fig:MRK463E_polar}. The coordinates of the nucleus, according to our polarization map, are : 13h56m02.91s, 18$^\circ$22'16.6''. A second estimation of the nucleus location, deduced from the centrosymmetric pattern of the polarization vectors onto the Northern polar outflow and inspired from the method of \citet{Kishimoto1999} agrees with this estimation and is presented in Appendix \ref{App:center}.

In polarized flux density, a bottleneck-like feature appears between the two hot spots. This apparent bridge between the extended clumps in the NLR could come from scattering of the UV continuum from the nuclei on material in the parsec-scale radio jet at position angle $~180^\circ$ \citep{Mazzarella1991,Kukula1999}. This could be investigated by correlating the present NUV polarization map to high resolution radio maps. Just outside the conical wind, $\sim1.5$~kpc East of the southern hot spot, a spherical clump (circled in blue in Fig. \ref{Fig:MRK463E_FOC_I} and \ref{Fig:MRK463E_FOC_PF_P}) appears in both total and polarized flux density with $P=19.8\pm4.9\%$ at $\Psi=76.2\pm7.1^\circ$. Two other spherical clumps (circled in yellow in Fig. \ref{Fig:MRK463E_FOC_I}) are visible in total intensity and polarization East of the Northern end of the polar wind, lying at $\sim2.8$~kpc and $\sim3.5$~kpc East of the major clumps lying North of the NLR cone, with $P=37.4\pm13.8\%$ at $\Psi=148.4\pm10.4^\circ$ and $P=11.4\pm5.9\%$ at $\Psi=56.5\pm14.8^\circ$ respectively. As both values are below $3\sigma_P$, it only hints for a possible polarization of the clumps, so it has to be taken with caution. South of the wind, two small knots (circled in green in Fig. \ref{Fig:MRK463E_FOC_I} and \ref{Fig:MRK463E_FOC_PF_P}) appear in polarization, at $\sim0.6$~kpc South-West and $\sim1$~kpc South-East of the vertex of the wind cone with $P=17.0\pm3.3\%$ at $\Psi=126.2\pm5.5^\circ$ and $P=24.4\pm4.3\%$ at $\Psi=77.0\pm5.1^\circ$ respectively.

All those features can be seen on the polarization degree map in the right-hand part of Fig.~\ref{Fig:MRK463E_FOC_PF_P}. This map reveals that the polarization inside of the NLR does not exceed $20\%$, dropping to zero around the center of the bottleneck structure. North of the NLR, the clumps show a polarization degree above $30\%$, increasing up to almost $50\%$ as they get further away from the first two arcseconds around the supposed position of the SMBH.

The centro-symmetric polarization pattern inside of the polar wind have an orientation that is coherent with the scattering of UV continuum coming from an unresolved source South of the cone. Such centro-symmetric pattern was also observed on IC~5063 and NGC~1068 in the previous paper of this series \citep{Barnouin2023} and reproduced the results of previous imaging polarimetry of NGC~1068, pin-pointing the location of the nucleus thanks to this circular pattern \citep{Capetti1995,Kishimoto1999}. We also note that the polarization angle of the escaping clumps North-East of the wind, also hint towards the same unresolved source that is South of the cone. The polarization angle pattern clearly proves the intuition of \cite{Uomoto1993}, who suggested that the base of the extended polar wind of Mrk~463E is acting like a mirror, allowing the observation of the hidden broad emission lines in polarized flux density.


\subsection{Comparison with previous polarimetric analysis}
\label{PreviousAnalysis}
Integrating over the whole HST/FOC FoV yields a total flux density of $F^\text{int}_\lambda = (17.81 \pm 0.02) \times 10^{-16}$ ergs cm$^{-2}$ s$^{-1}$ \AA$^{-1}$, with a polarization degree $P^\text{int} = 8.8\pm0.2\%$ and a global polarization position angle $\Psi^\text{int}= 88.3\pm0.5^\circ$. Cropping the observation to remove Mrk~463W, as it can be seen in Fig.~\ref{Fig:MRK463E_FOC_PF_P} and \ref{Fig:MRK463E_knots} yields $F^\text{int}_\lambda = (16.18\pm0.01) \times 10^{-16}$ ergs cm$^{-2}$ s$^{-1}$ \AA$^{-1}$, $P^\text{int}= 10.0\pm0.2\%$ and $\Psi^\text{int} = 87.9\pm0.4^\circ$. This cropped polarization map, only showing Mrk~463E, will be used in the following analysis. Finally, integrating only the contribution from the polar wind (including the escaping clumps) yields a polarization degree $P^\text{int} = 12.5\pm0.2\%$ at position angle $\Psi^\text{int} = 87.5\pm0.3^\circ$, clearly indicating that most of the polarized flux density comes from reprocessing in the winds. In all cases, the polarization angle is perpendicular to the $\sim4^\circ$ radio position angle measured by \citet{Unger1986} and \citet{Neff1988}, as expected for a type-2 AGN.

We can compare our HST/FOC map to previous observations. It is a fairly common target and a plethora of spectropolarimetric data exists. A direct flux-to-flux comparison with \citet{Watanabe2003} validates the flux density measured by HST: the authors measured a total flux density of $6.8 \times 10^{-15}$ ergs cm$^{-2}$ s$^{-1}$ \AA$^{-1}$ at 5400~\AA\ in a $3''$ extraction window on 2001, February 21. The slight difference between the two fluxes is easily explained due to the different wavelength windows (the total flux spectrum of Mrk~463 rising to the blue) and also to the seeing conditions on Earth ($1.3-1.4''$) and (the lack-of) in space. \citet{Watanabe2003} acquired polarized spectra of Mrk~463E from 4600~\AA\, to 25\,000~\AA\, and their polarization degree and angle ($10.4\%$ and $82^\circ$) at 5400~\AA\ agree well with our measurements at 3404~\AA. This indicates that electron scattering could be at work here, since the polarization is roughly wavelength-independent, as expected for electron scattering versus dust scattering that would generally increase with decreasing wavelength \citep{Miller1991}, but the lack of error bars on \citet{Watanabe2003}'s measurements hamper a strong conclusion.

\cite{Young1996} obtained optical and near-IR broad-band polarimetric observations of Mrk~463E using the 3.8m United Kingdom Infrared Telescope (UKIRT) and a $5.0''$ aperture. They report  $P = 6.48\pm0.5\%$ at $\Psi = 86\pm2^\circ$ in the U-band and $P = 5.78\pm0.28\%$ at $\Psi = 88\pm1^\circ$ in the V-band, both in agreement with our full FoV polarization at 3404~\AA\ that benefited from less unpolarized light contamination. Mrk~463E was also observed both in imaging polarimetry and spectropolarimetry with the 3m Shane telescope at Lick observatory. The spectropolarimetric analysis was performed through a North-South oriented, $2.4''$ wide and $10''$ long slit (see Fig.~30 in \citealt{Tran1995b}) and presented in \cite{Miller1990} and \cite{Tran1995a}. They obtain a continuum polarization of $P = 7.6\%$ at $\Psi = 85^\circ$ in the 3200 -- 7400~\AA\ (they also report a broad line polarization of $P = 10\%$ at $\Psi = 84^\circ$). We simulate such slit by integrating the I, Q and U Stokes fluxes inside a $2.4''$ wide and $7''$ long slit (limited by the $7'' \times 7''$ FoV of the HST/FOC) oriented North-South and centered on the [O III] maximum flux, see Fig.~\ref{Fig:MRK463E_FOC_I}. We report an integrated flux density around 3404~\AA\ of $F^\text{int}_\lambda = (12.57 \pm 0.01) \times 10^{-16}$ ergs cm$^{-2}$ s$^{-1}$ \AA$^{-1}$ and a polarization degree of $P = 10.8\pm0.2\%$ at position angle $\Psi = 87.6\pm0.4^\circ$, also compatible with the previously reported optical polarization.

\subsection{Region specific polarization analysis}
\label{RegionAnalysis}
\begin{figure}
    \centering
    \includegraphics[width=1.\linewidth]{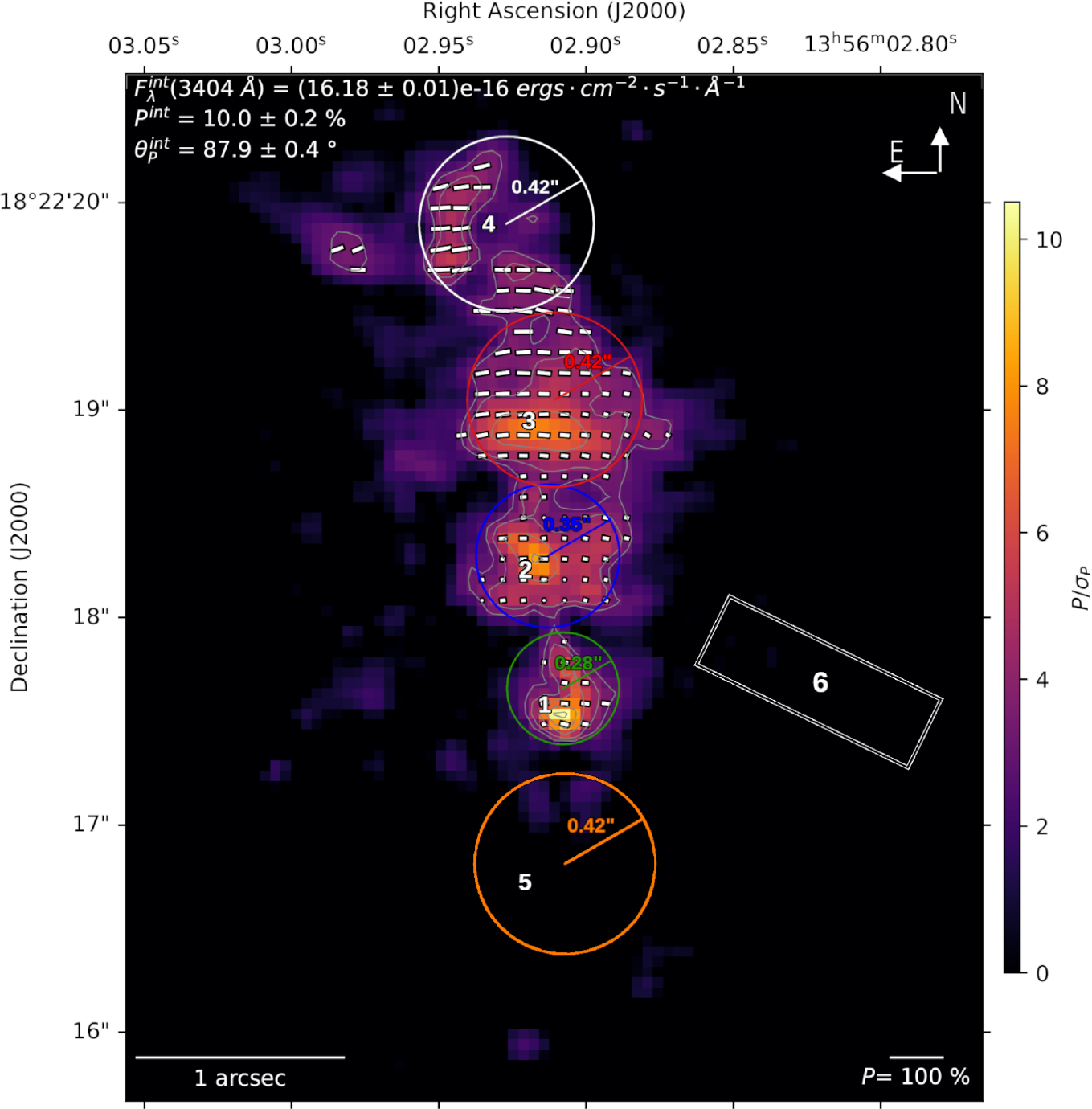}
    \caption{Knots in the North polar wind. For better visibility we only display every other vector. The 6$^\text{th}$ region corresponds to a simulated slit aperture of $0.35'' \times 1.15''$ along the the stream-like feature that can be observed in Fig. \ref{Fig:MRK463E_FOC_I} and which is discussed in Sect.~\ref{Filament}.}
    \label{Fig:MRK463E_knots}
\end{figure}

The unmatched imaging resolution of the HST/FOC in the NUV band allows to perform region specific polarimetric studies. In the following, we use simulated apertures and slits to obtain integrated polarization characteristics on selected regions. Regions for aperture integration are shown in Fig. \ref{Fig:MRK463E_knots} on top of the S/N map in polarization degree, with the contribution from Mrk~463W cropped-out. This particular map allow us to select region of high polarization degree, based on the most significant polarization detection. We report the integrated values on each region for flux density, polarization degree and angle in Tab. \ref{Tab:knots_pola}.

Regions 1, 2 and 3 correspond to clumpy structures located inside of the northern polar wind showing high polarized fluxes (see Fig. \ref{Fig:MRK463E_FOC_PF_P}). As we probe clumps further away from the nucleus, we move along the polar axis and find that the clumps conserve a polarization angle that is coherent with perpendicular scattering. The slightly different polarization angle $\Psi^2 \simeq 87.6^\circ$ of region 2 with respect to regions 1 and 3 can easily be explained by the fact that the center of the polarized flux knot is displaced East of the polar axis.

Region 4 was selected to focus on the escaping clumps North-East of the wind. The integrated polarization shows a polarization angle of $\Psi^4 \simeq 92.5^\circ$, a few degrees higher than region 1 to 3 in the cone, but still consistent with diffusion of the continuum from the nucleus. These clouds are also characterized by a strong polarization degree of $P^4 \simeq 34\%$. This could be due to a variety of reasons, such as less perturbed medium than in the cone, less obscuration along the line-of-sight, higher albedo, etc., which are indistinguishable in this context.

Region 5 corresponds to the location of the hidden southern polar cone. While it is not clear at this stage if the wind is simply absent or obscured (see Sect.~\ref{WFPCAnalysis} for further discussion), we wanted to check if polarization could be detected here. We find a small but statistically significant ($\sim5.5\sigma_P$) polarization degree associated with a polarization angle perpendicular to the jet direction. It indicates that there is a scattering region, likely in the form of a wind (or at least clumps), but the total flux signal seems very attenuated, probably by dust absorption.

The stream-like feature going from Mrk~463E toward Mrk~463W and visible in total intensity only is the focus of region 6. We simulated a $0.35''$ wide and $1.15''$ long slit, oriented $67^\circ$ East of North, to integrate the Stokes fluxes especially along the hypothetical stream. Unfortunately, the detected polarization is below $3\sigma_P$ and nothing strong can be claimed here without further observation. Nevertheless, if the structure is real, we have hints for a polarization degree $P^6 = 3.8\pm1.9\%$ at an electric polarization angle $\Psi^6 = 133.4\pm14.3^\circ$ (with 89\% confidence level for two degrees of freedom on the Stokes Q and U fluxes). We will discuss these results in Sect.~\ref{Filament}.

\begin{table}
    \centering\small
    \resizebox{\linewidth}{!}{%
        \begin{tabular}{c|c|c|c}
            Region  & Flux density ($10^{-17} \text{ergs}\;\text{cm}^{-2}\;\text{s}^{-1}\;\text{\AA}^{-1}$) & P (\%) & PA ($^\circ$)\\
            \hline
            Full FoV    & $178.10 \pm 0.20$ & $8.8 \pm 0.2$ & $88.3 \pm 0.5$ \\
            Cropped FoV & $161.80 \pm 0.10$ & $10.0 \pm 0.2$    & $87.9 \pm 0.4$ \\
            Cone+clumps & $105.70 \pm 0.10$ & $12.5 \pm 0.2$    & $87.5 \pm 0.3$ \\
            $2.4'' \times 7''$ slit  & $125.70 \pm 0.10$ & $10.8 \pm 0.2$    & $87.6 \pm 0.4$ \\
            1   & $14.77 \pm 0.03$  & $12.8 \pm 0.4$    & $85.3 \pm 0.8$ \\
            2   & $35.25 \pm 0.05$  & $11.0 \pm 0.3$    & $87.6 \pm 0.6$ \\
            3   & $19.16 \pm 0.03$  & $17.6 \pm 0.3$    & $84.7 \pm 0.5$ \\
            4   & $3.30 \pm 0.02$   & $33.9 \pm 0.9$    & $92.5 \pm 0.8$ \\
            5   & $2.21 \pm 0.02$   & $6.0 \pm 1.1$ & $99.3 \pm 5.3$ \\
            6   & $0.96 \pm 0.01$   & $3.8 \pm 1.9$ & $133.4 \pm 14.3$
        \end{tabular}%
    }
    \caption{Integrated flux density and polarization over several regions of interest (see Fig.~\ref{Fig:MRK463E_knots}). Reported polarization degree are corrected for bias \citep{Simmons1985}}
    \label{Tab:knots_pola}
\end{table}

\section{Multi-wavelength Analysis}
\label{MultiwavAnalysis}
\subsection{Correlation with HST/WFPC2 near-IR observation}
\label{WFPCAnalysis}
The Mrk~463 system was observed using the Wide Field Planetary Camera 2 (WFPC2) onboard HST in November 1995 as part of a sample of ten near infrared quasars (Proposal ID 5982). The binary system was observed through the F814W filter, centered around 7996~\AA\ with a bandwidth of 646~\AA. The results of this observation were published in \cite{Sanders1996}. While the WFPC2 has a spatial resolution similar to the FOC (pixels of size $0.0455'' \times 0.0455''$), it benefits from a much wider FoV of $36'' \times 36''$. We thus decided to take advantage of the WFPC2 observation of Mrk~463E in order to better understand the geometric configuration of matter around the nucleus, as well as its composition.

\begin{figure*}
    \centering
    \includegraphics[width=1.\textwidth]{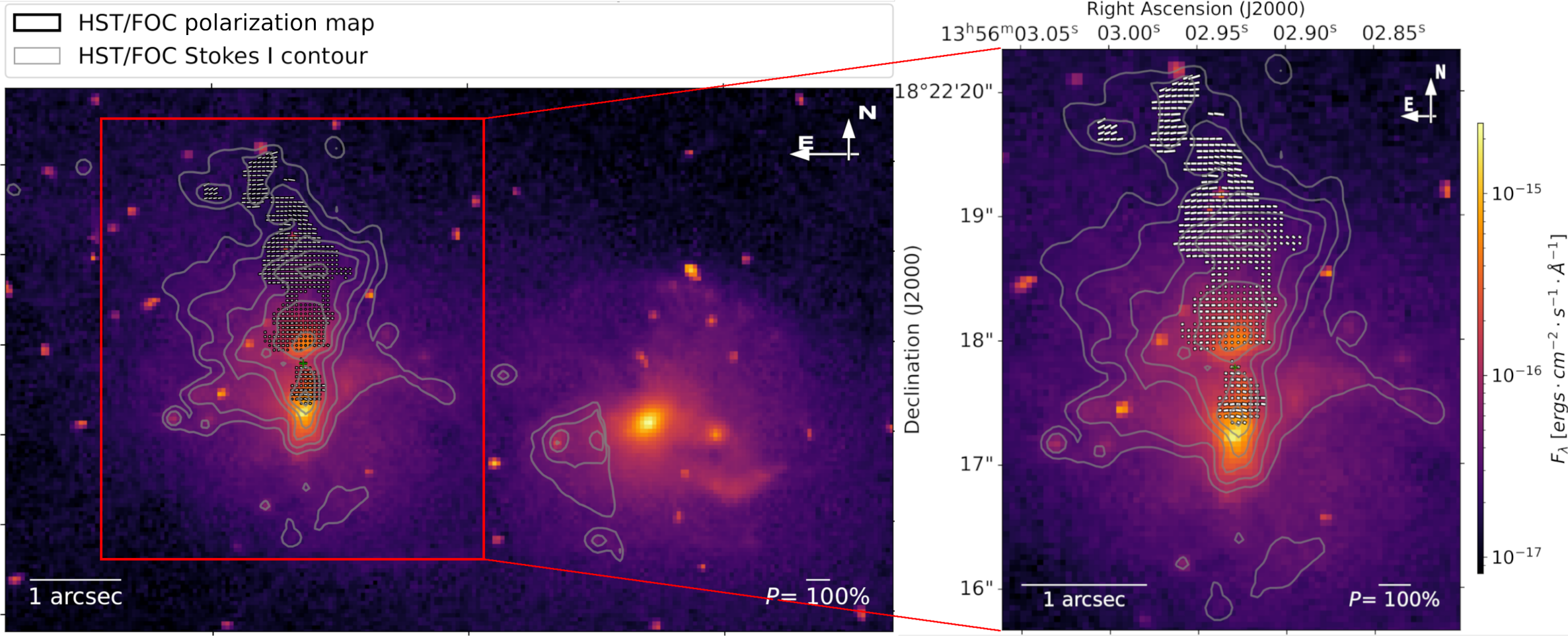}
    \caption{WFPC2 observation of Mrk~463, overlaid with the polarization vectors obtained from the FOC observation analysis. Both NIR and NUV maps were aligned on similar looking features around the nucleus of Mrk~463E.}
    \label{Fig:MRK463E_WFPC2}
\end{figure*}

We downloaded and reduced the WFPC2 data, and overlaid in Fig.~\ref{Fig:MRK463E_WFPC2} the FOC total intensity (gray contours) and polarization (white vectors) maps on top of the near-infrared total intensity map taken with the F814W filter. Both maps were aligned on similar-looking punctual and extended features, as it can be seen on the zoomed box in the right-hand part of Fig.~\ref{Fig:MRK463E_WFPC2}. The wider FoV of WFPC2 allowed to observe both nuclei at once and better assessed the contribution of the western object to the FOC image. From this simple overlay we can see that the Mrk~463E maximum intensity regions in NUV and NIR do not overlap, with the NIR peak being just south of the NUV peak. We also note that the stream-like feature seen West of Mrk~463E and extending in the general direction of Mrk~463W in the NUV intensity map also appear in the NIR intensity map, reinforcing the probability of being a real structure and not an observational artifact of the FOC.

\begin{figure}
    \centering
    \includegraphics[width=1.\linewidth,trim={0 0 0 12mm},clip]{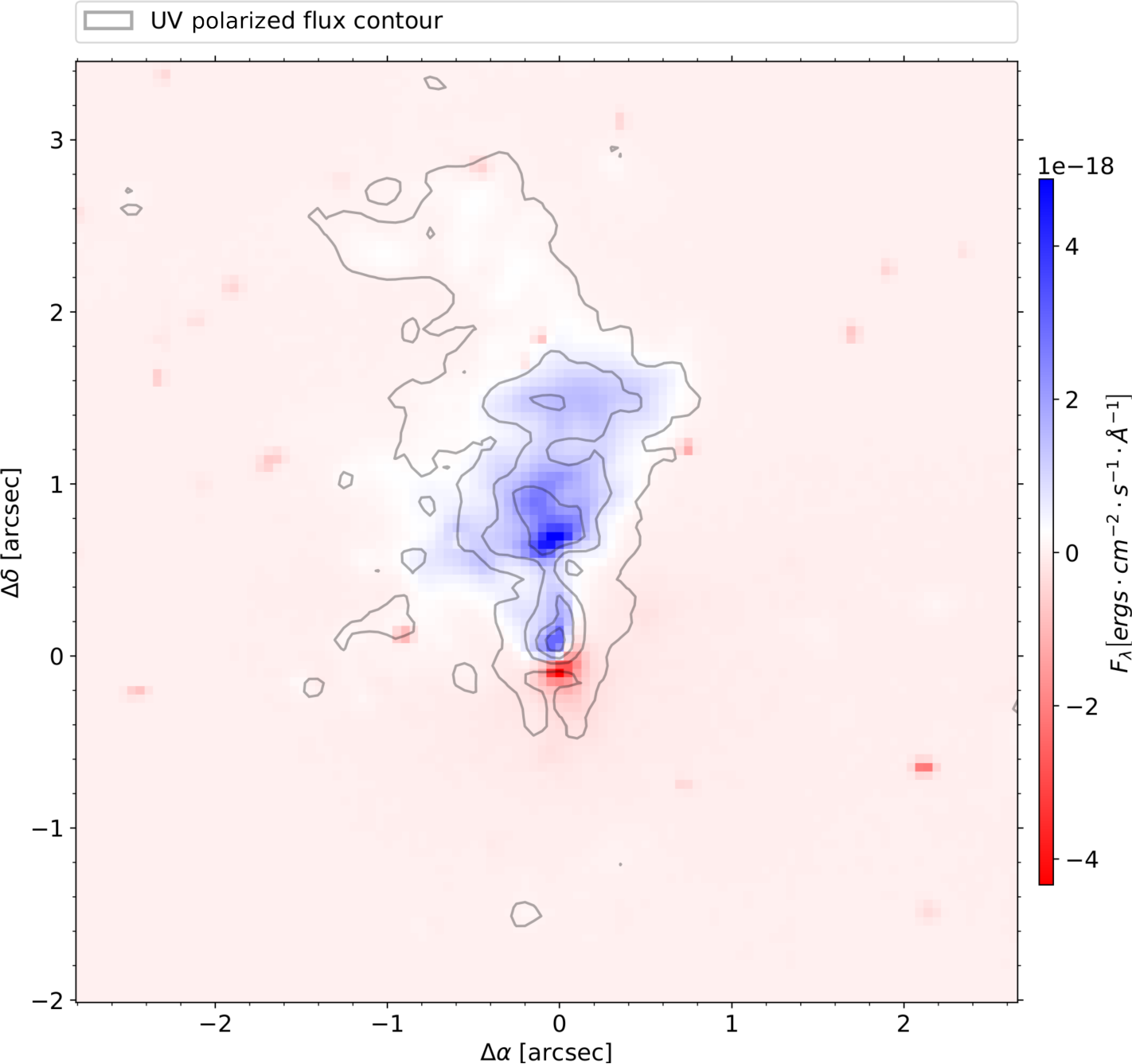}
    \caption{HST/WFPC2 image at 7996~\AA\ subtracted to the HST/FOC image at 3404~\AA. Both maps were binned to $0.05''$ pixel size and aligned on similar looking features. The contours display the level of NUV polarized flux density, correlated with the polar wind of the AGN.}
    \label{Fig:MRK463E_reddening}
\end{figure}

Because the NUV and NIR maps have the same pixel size and have been properly aligned, it is possible to subtract one from the other. This is what we did to obtain the excess NUV map, visible in Fig.~\ref{Fig:MRK463E_reddening}. It is a worthwhile exercise as the NUV excess can be used to determine whether the UV light could be attributed to recent star formation (which, in turn, can be linked to evolutionary processes in the host galaxy) or a direct consequence of the AGN activity \citep{Fabian1989}. Our map clearly shows that the latter explanation is correct in the case of Mrk~463E: the NUV excess perfectly matches the polar wind geometry. It is thus a solid evidence advocating for reprocessing of the the UV-NIR continuum emission from the illuminating source (i.e., the nucleus) onto the Northern winds. Additionally, it indicates that the Northern winds are located in front of the host's galactic plane, in a similar fashion than NGC~1068 \citep{Kishimoto1999}. The Southern winds are very likely obscured by the dusty content of the host galaxy. It is interesting to note that there is a deficit of UV photons precisely at the base of the northern wind. Correlating this NUV-excess map to radio maps of the parsec-scale jets could provide additional information to constrain more precisely the location of the nucleus.


\subsection{Correlation with Chandra 2-8~keV observation}
\label{ChandraAnalysis}
In the high energy domain, where photon fluxes are weaker, it is not easy to acquire high resolution images. It is detrimental because arcsecond X-ray imaging could provide a wealth of information about black hole growth with time, through mergers (likewise like Mrk~463E) or AGN fueling \citep{Mushotzky2019}. Luckily, the Chandra X-ray Observatory was pointed toward this system in June 2004 (ObsId 4913) at took full advantage of its Advanced CCD Imaging Spectrometer (ACIS). The observation allowed for the first, firm identification of both nuclei as Seyfert-2s \citep{Bianchi2008}.

\begin{figure}
    \centering
    \includegraphics[width=1.\linewidth,trim={0 0 0 21mm},clip]{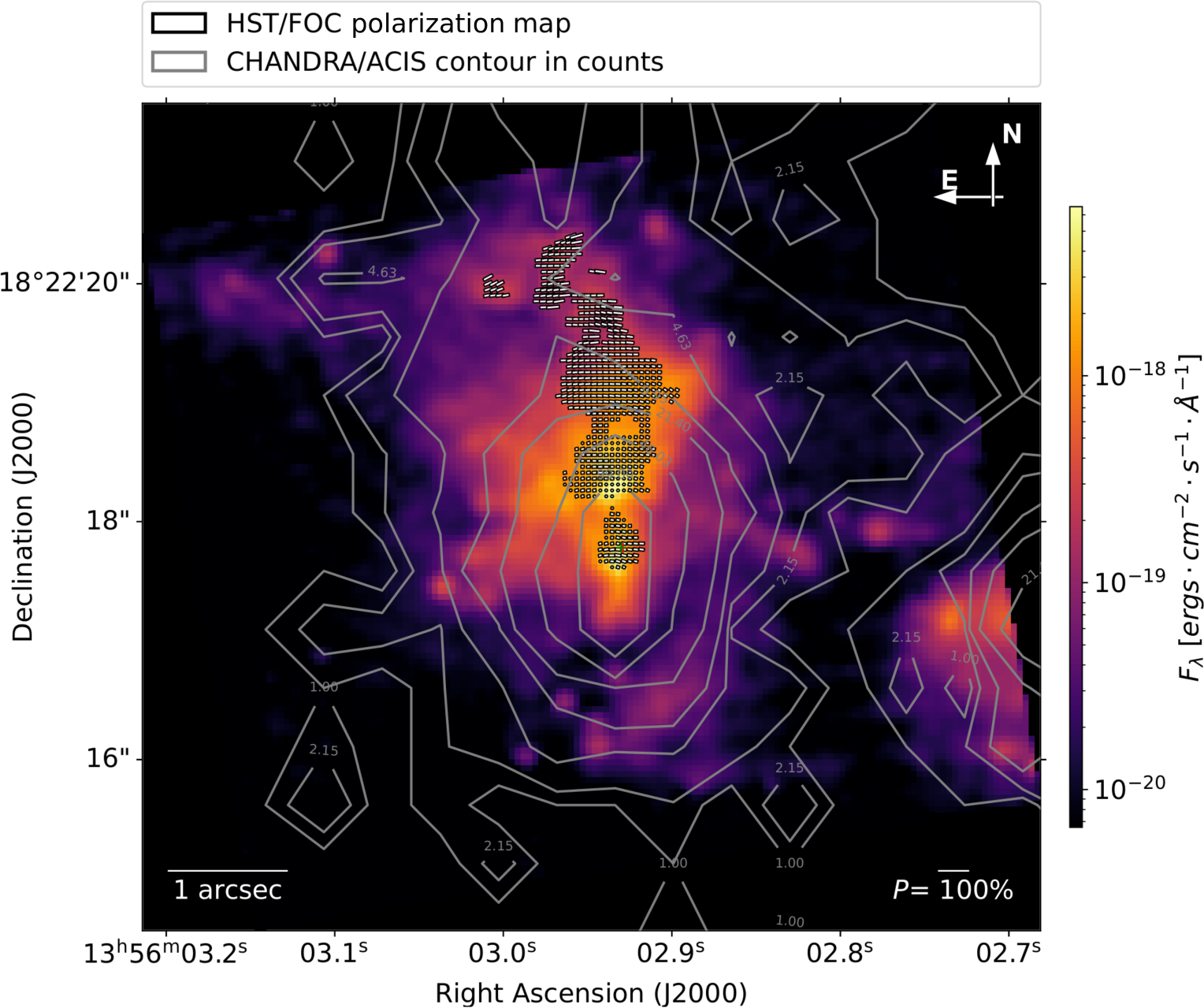}
    \caption{Chandra observation of Mrk~463E, shown with contours overlaid on the FOC observation analysis, showing both the near-UV flux density and polarization map. The contours corresponds to every tenth percentile of the maximum Chandra counts.}
    \label{Fig:MRK463E_Chandra}
\end{figure}

Similarly to the WFPC2 case, we downloaded and reduced the data in order to overlay the soft X-ray emission onto the FOC map. The result of such exercise is presented in Fig.~\ref{Fig:MRK463E_Chandra}. The alignment was more complex as the spatial resolution of the Chandra map ($0.492'' \times 0.492''$) is one order of magnitude "worse" than that of the HST/FOC. We used the fact that both AGNs (Mrk~463E and Mrk~463W) were observed by Chandra to align the two objects with the NUV map. Ultimately, we can see that the peak of X-ray emission is consistent with the peak of NUV flux in the first arcsecond around the hidden nucleus, but subtle details are elusive. We also see that, despite the X-rays being more extended than the NUV emission, the two nuclei are well separated on the Chandra map, confirming the non-contamination of the Mrk~463E HST/FOC image from the western nucleus.

\section{Discussion}
\label{Discussion}
\subsection{A streamer connecting Mrk~463E and Mrk~463W?}
\label{Filament}
Thanks to the sharp contrast of the total intensity of the HST/FOC NUV map against the diffuse background, we observed an elongated faint source extending from the northern wind of Mrk~463E towards the South-West, in the general direction of Mrk~463W (see Fig.~\ref{Fig:MRK463E_FOC_I}). This stream has an NIR counterpart in the WFPC2 observation, slightly less contrasted than in the FOC map. This NIR/NUV comparison allows us to confirm that the stream has a pronounced curvature that connects to the base of the northern wind of Mrk~463E (see Fig.~\ref{Fig:MRK463E_WFPC2}). For this reason, one might think that it is a stream of gas and dust linked to the past encounter between Mrk~463E and Mrk~463W, rather than a filament related to the AGN winds.

In order to analyze this structure more closely, we integrated the polarization of such stream on the HST/FOC image using a simulated slit of $0.35'' \times 1.15''$ at a $67^\circ$ angle, so that it gets a maximum of flux from the stream itself and as little pollution as possible from the AGN wind and the background. The polarization detected at 89\% confidence level, reported in Sect.~\ref{RegionAnalysis} and Tab.~\ref{Tab:knots_pola}, hints for an electric vector polarization angle of $\Psi^6 \sim 133^\circ$. Such polarization angle is inconsistent with the hypothesis that the AGN nuclei (either Mrk~463E or Mrk~463W) are the primary sources of radiation scattering onto the stream. For example, for a scattering target located at the center of our extraction region, the line orthogonal to the direction towards the hidden nucleus of Mrk~463E would correspond to a polarization angle of $41^\circ$.

The low confidence on this measure allows only for a speculative discussion. The polarization from this potential stream is thus uncorrelated to the AGN (otherwise, it would have probably appeared on the NUV excess map too). The fact that the magnetic vector, orthogonal to the measured electric vector polarization angle, roughly aligns with the stream direction suggests that the origin of the polarization could very well be dichroic absorption from Mrk~463E host starlight (or within the stream itself) in the foreground dusty stream (we recall that Mrk~463E is situated further away than Mrk~463W from us). The integrated polarization angle could then highlight the large kpc-scale ordered magnetic field in the stream. Indeed, such gas streamers with highly compressed magnetic fields have been found in interacting galaxies both in radio \citep{Drzazga2011} and far-IR \citep{LopezRodriguez2023}. Despite a measurement at 89\% confidence level only, the number of evidences pointing toward a real stream connecting both nuclei are intriguing. New IR and radio observations are mandatory to explore this feature in more depth.

\subsection{What do the winds tell us?}
\label{Winds}

\begin{figure}
    \centering
    \includegraphics[width=1.\linewidth]{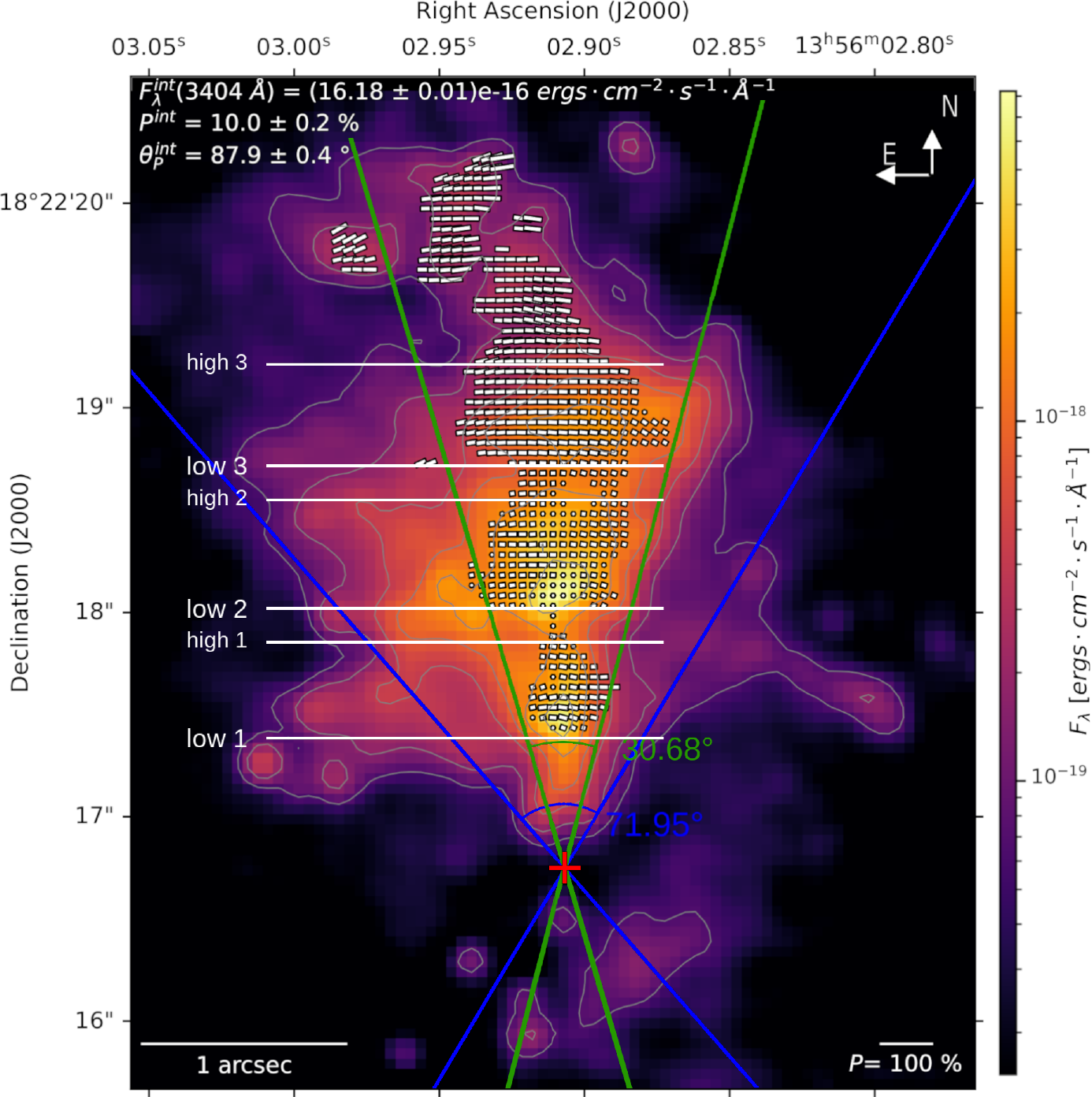}
    \caption{Total intensity NUV map from the HST/FOC cropped around Mrk~463E northern winds. The blue and green lines represent the total opening angle for the wind cones as seen in total intensity ($71.95^\circ$) and polarized flux ($30.68^\circ$), respectively. The red cross shows our estimation for the nucleus location.}
    \label{Fig:MRK463E_polar}
\end{figure}

The polarization map allows us to study the two components of the polar winds -- the NLR and the ionization cone, the former being visible in total intensity and the later in polarized flux density. Both are well geometrically delimited against the background emission, so we can measure their half opening angle in Fig.~\ref{Fig:MRK463E_polar}. The NLR shows an half opening angle of $\sim36^\circ$ with respect to the jet axis of the system, which is in agreement with the first results from \cite{Tremonti1996}, while the ionization cone, revealed in polarized flux density, only has a half opening angle of $\sim15^\circ$. Inside the ionized cone, the polarization angle stays uniform at $\sim90^\circ$ with very small dispersion. On average, high polarization degrees are well correlated with peaks of total intensity, which can be related to higher cloud densities. We report almost no polarization around the bottleneck between the two NUV hot spots (the region between the white lines noted "high~1" and "low~2" on Fig.~\ref{Fig:MRK463E_polar}). This could result from kinematics constraints inside the wind, sudden lack of scatter material in this region or from a variation in the past ejection activity of Mrk~463E. In the latter case, continuum photons may scatter on the radio jet between two periods of mass ejection, which would explain the bridge-like feature in both total and polarized flux densities. High resolution imaging of the radio jets is required to confirm this hypothesis.

From the excess map in Fig.~\ref{Fig:MRK463E_reddening}, we now know that the northern wind is above the host galaxy's plan, a result that is confirmed by the observation of negative velocity with respect to the systemic velocity obtained from mapping the [OIII] 5007~\AA\ line emission, see Fig. 12 (left) in \cite{Treister2018}. However, the inclination of the winds is unknown. We may still try to infer the global inclination of the AGN thanks to polarimetry. If we assume that the NUV radiation we measured consists of only radiation scattered by free electrons and no other diluting component exists, we can then deduce the scattering angle from the nuclei towards the line of sight within the framework of single scattering:
\begin{equation}
    P = \frac{1-\mu^2}{1+\mu^2},
\end{equation}
where $\mu = \cos{\theta}$ and $\theta$ is the scattering angle. This relation works well if the scattering target size is small enough compared with the distance to the illuminating source \citep{Kishimoto1999}. Considering only pixels with $\left[S/N\right]_P \geq 5$ in region 3 (see Fig.~\ref{Fig:MRK463E_knots}), we find an averaged $P$ of $20.8 \pm 0.6\%$, which translates into two solutions for $\theta$ : $\theta_1 = 144.1 \pm 0.5^{\circ}$ and $\theta_2 = 35.9 \pm 0.5^{\circ}$ (uncertainties on $\theta$ were propagated from the previous formula). As a result, we can conclude that Mrk~463E is probably inclined by about $90^{\circ} - 35.9^{\circ} = 54.1^{\circ}$ (rounded to a value of $55^{\circ}$ for simplicity) with respect to the jet axis, well within the expected range of inclinations for a type-2 object \citep{Marin2014,Marin2016}.

Additionally, the shape of the winds, revealed both in total intensity and polarimetry, shows three regions (regions 1, 2 and 3 in Fig.~\ref{Fig:MRK463E_knots}) which resemble compact clusters of material illuminated by the hidden nuclei and separated by regions of flux under-density. This could correspond to periods of massive polar ejection spaced by periods of calmer nuclear activity. Those gaps between the different clumps of matter are particularly visible in the polarized flux map between region 1 and 2 (see Fig.~\ref{Fig:MRK463E_FOC_PF_P}). Now that we have a tentative value for the inclination of the AGN, we can estimate the duration of the activity periods that lead to massive polar ejection. Based on our analysis, we suppose an inclination angle towards the observer of $\theta_\text{incl} = 55^\circ$ and a uniform wind velocity $V = 350$ km s$^{-1}$ \citep{Treister2018}. Given the hypothesis that the origin of the wind is the nucleus \citep{Joh2021}, whose location is determined by the cone apexes in Fig. \ref{Fig:MRK463E_polar}, we can deduce the propagation time:
\begin{equation}
    T = \frac{a_{z} \cdot A^\text{obs}}{V \cdot \cos{\theta_\text{incl}}} \simeq 5.0455 \cdot A^\text{obs} \; \text{Myr},
    \label{Eq:Winds_propag}
\end{equation}
where $a_{z}$ is the angular to linear size conversion factor for Mrk~463E at $z = 0.05132$ in standard $\Lambda$CDM cosmology, and $A^\text{obs}$ is the observed angular size on the HST/FOC map in arcsec. We report the results in Tab. \ref{Tab:Winds_time}. From our analysis, we deduce that the clouds comprised in regions 1, 2 and 3 in Fig. \ref{Fig:MRK463E_knots} were ejected from the nuclei about 4.3, 7.6 and 11.2 Myr ago respectively. These three clouds display similar ejection duty cycles, with massive ejection lasting for 2.3 Myr, with 1.1 Myr of quiet state in-between. Such orders of magnitude are representative of typical AGN duty cycles such as estimated by, e.g., \citet{Turner2018}, \citet{Biava2021} or \citet{Maccagni2021}.

\begin{table}
    \centering
    \begin{tabular}{c|c|c}
        Region & Angular size (arcsec) & Propagation time (Myr)\\
        \hline
        lower 1 & 0.64 & 3.21\\
        higher 1 & 1.08 & 5.43\\
        lower 2 & 1.27 & 6.38\\
        higher 2 & 1.75 & 8.84\\
        lower 3 & 2.00 & 10.08\\
        higher 3 & 2.44 & 12.30
    \end{tabular}
    \caption{Angular distance from the nucleus to the inner and outer edges of each cloud and its corresponding propagation time, computed from Eq.\ref{Eq:Winds_propag}. Each distance is taken from the apex of the cone to the associated region indicated in Fig.\ref{Fig:MRK463E_polar}.}
    \label{Tab:Winds_time}
\end{table}

\section{Conclusion}
\label{Conclusion}
In this second paper of our series, we made use of the general pipeline for reduction of the HST/FOC archival data presented in \cite{Barnouin2023} to perform the analysis of a previously unpublished dataset. The archived observation of Mrk~463E revealed to be a true forgotten treasure and a great illustration of the strength of high spatial resolution NUV polarimetry. The strong polarized flux from the source made it possible to extensively study the NLR and granted us an improved estimation of the location of the nucleus, further South than what was previously suggested (estimated coordinates: 13h56m02.91s, 18$^\circ$22'16.6'').

This high angular resolution polarization map also gave a tight estimation of the AGN inclination angle, which we found to be about $55^\circ$. We also measured the polar wind's half opening angle between $15^\circ$ (in polarization) and $35^\circ$ (in total intensity). The observed polarization is in good agreement with previous spectropolarimetric studies of Mrk~463E and calls for additional IR studies of the stream-like feature (the gas streamer) between the two interacting nuclei of the system. This analysis can further be completed with high resolution radio imaging of the jets to better understand their interaction with the observed NLR.

\begin{acknowledgements}
TB and FM would like to acknowledge the support of the CNRS, the University of Strasbourg, the PNHE and PNCG. This work was supported by the "Programme National des Hautes Énergies" (PNHE) and the "Programme National de Cosmologie et Galaxies (PNCG)" of CNRS/INSU co-funded by CNRS/IN2P3, CNRS/INP, CEA and CNES. E.L.-R. is supported by the NASA/DLR Stratospheric Observatory for Infrared Astronomy (SOFIA) under the 08\_0012 Program. SOFIA is jointly operated by the Universities Space Research Association,Inc.(USRA), under NASA contract NNA17BF53C, and the Deutsches SOFIA Institut (DSI) under DLR contract 50OK0901 to the University of Stuttgart. E.L.-R. is supported by the NASA Astrophysics Decadal Survey Precursor Science (ADSPS) Program (NNH22ZDA001N-ADSPS) with ID 22-ADSPS22-0009 and agreement number 80NSSC23K1585.
\end{acknowledgements}

\bibliographystyle{aa}
\bibliography{biblio}

\appendix

\section{Confidence level for the polarization degree.}
\label{App:Conf}

We propose a two parameters confidence level for the polarization degree that differ from the $4-\sigma_P$ detection level proposed by \citet{Simmons1985}. This method relies solely on the scalar Stokes $Q$ and $U$ fluxes, computed in units of counts per second throughout the whole pipeline. These can be modeled along with gaussian noise.
As per definition of the Stokes fluxes, we compute the polarization degree as $P = \sqrt{q^2+u^2}$, with $q=Q/I$ and $u=U/I$ the normalized Stokes $Q$ and $U$ fluxes. We define the confidence at which we detect the polarization on the $\chi^2$ statistic on the normalized Stokes $Q$ and $U$ fluxes with their associated normalized uncertainties:
\begin{equation}
    \chi^2 = \frac{q^2}{{\sigma_q}^2}+\frac{u^2}{{\sigma_u}^2}
    \label{eq:chidetect}
\end{equation}
The confidence level is computed on the cumulative distributive function for $\chi^2$ with two parameters of interest
\begin{equation}
    CL = CDF(\chi^2, 2) = 1 - e^{-\frac{1}{2}\chi^2}
    \label{eq:chiconf}
\end{equation}

\begin{figure*}
    \centering\includegraphics[width=0.99\linewidth]{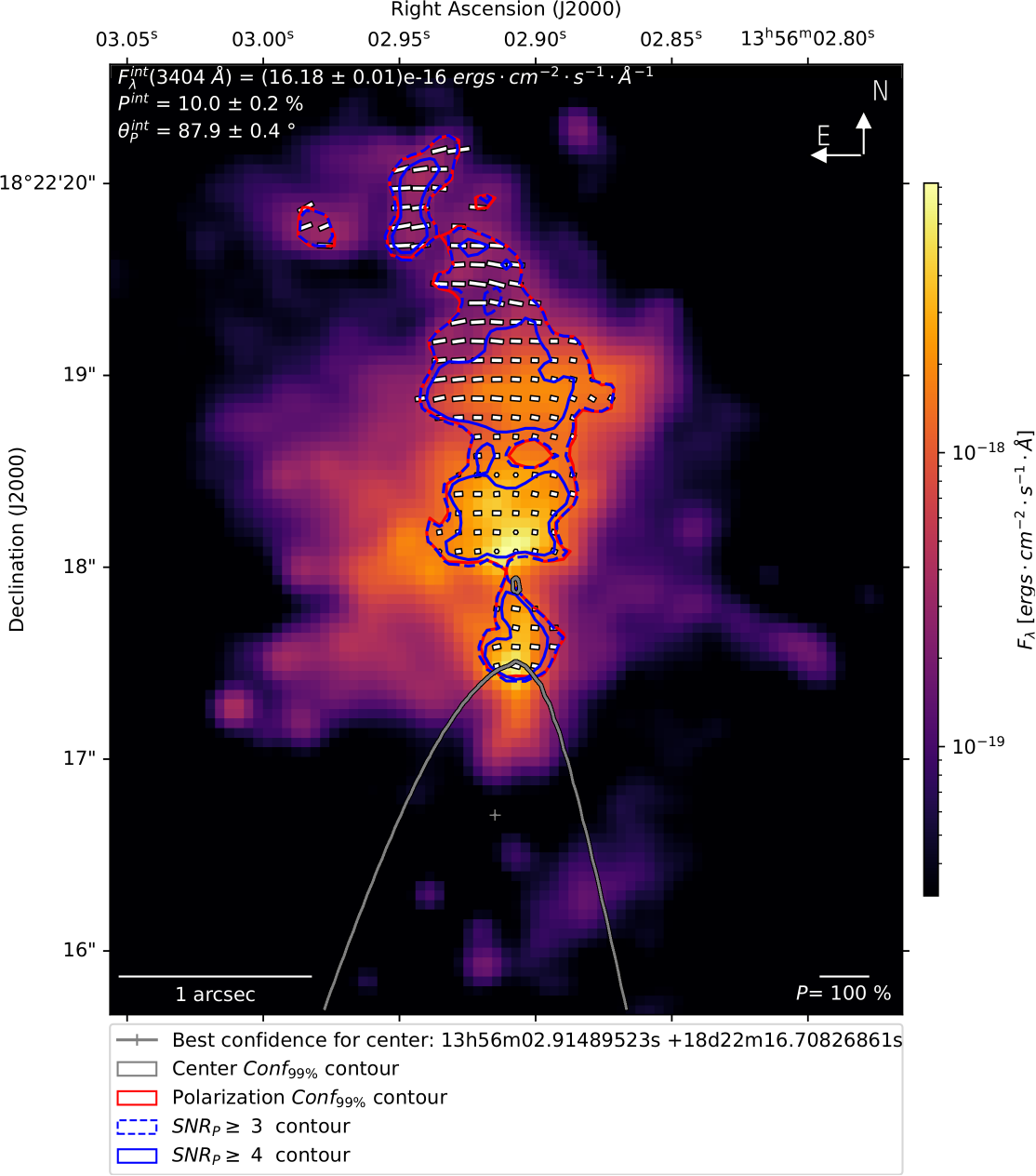}
    \caption{Comparison of the determination of a detection between levels of signal to noise ratio in debiased polarization degree in blue for $3-\sigma_P$ (dashed line) and $4-\sigma_P$ (solid line) with the Stokes Q and U fluxes confidence level defined in Eq. \ref{eq:chiconf} in red. The gray contour and position are the 99\% confidence level and best guess for the position of the center of emission by minimizing Eq. \ref{eq:chicenter}.}
    \label{fig:ConfCenter}
\end{figure*}

We compare the $3-\sigma_P$ and $4-\sigma_P$ detection levels on the flux density map of Mrk~463E in Fig. \ref{fig:ConfCenter}. In our case, the 99\% confidence level as defined in Eq. \ref{eq:chiconf} encompass a region less restrictive than the $4-\sigma_P$ level, but slightly more restrictive than the $3-\sigma_P$ level, in accordance with the suggestion by \citet{Simmons1985}.

\section{Localization of the emission center}
\label{App:center}
The single scattering of a photon from a point source onto an ionized polar outflow produces polarization with an electrical polarization angle perpendicular to the direction to the source. Based on the full uncertainties on the computed polarization angle, we evaluate the confidence with which the detected polarization is consistent with point-source scattering.
\begin{equation}
    \chi^2(x_c,y_c) = \sum_{i \in A} \frac{\left[\theta_\text{PA}(i)-\theta_\text{ideal}(i;x_c,y_c)\right]^2}{{\sigma_\text{PA}(i)}^2}
    \label{eq:chicenter}
\end{equation}
with $A$ a valid ($\ge$ 5 sigmas) subset of polarization vectors, $\theta_\text{PA}$ the observed polarization angle and $\sigma_\text{PA}$ its associated full uncertainty as defined in \citet[Eq 21]{Barnouin2023}.
For any vector $i$ at pixel coordinates $x_i,y_i$, the ideal centrosymmetric PA for a center of emission at pixel coordinates $x_c,y_c$ is:
\begin{equation}
    \theta_\text{ideal}(i;x_c,y_c) = \arctan\left(\frac{y_c-y_i}{x_c-x_i}\right)
\end{equation}
By minimizing the value of $\chi^2$, we find our best estimation for the location of the nucleus in the hypothesis of single scattering onto ionized medium, assuming for scattering on a single plane. We show the 99\% confidence level contour and the emission center that minimize Eq. \ref{eq:chicenter} with $A$ the set of vectors with a polarization confidence greater than 99\%. Both the contour and best estimate agree with the estimation from the wind cones openings in Sect. \ref{FOCmap} and Fig. \ref{Fig:MRK463E_polar}. We note that the 99\% contour does not exclude for a nucleus location as previously suggested in the literature \citep{Uomoto1993} but this hypothesis remains marginal.

\end{document}